\documentclass[12pt]{iopart}
\usepackage{times}
\usepackage{mathptmx}
\usepackage{graphicx}
\usepackage[dvips,colorlinks]{hyperref}

\begin{document}

\centerline { \bf SECOND-ORDER LINEAR DIFFERENTIAL EQUATIONS} 
\centerline{\bf }
\centerline{\bf ASSOCIATED TO A RICCATI EQUATION OF CONSTANT COEFFICIENTS}

\bigskip 

\centerline {H.C. Rosu$^1$\footnote{hcr@ipicyt.edu.mx}, O. Cornejo$^1$, M. Reyes$^2$, D. Jimenez$^2$} 

\bigskip

\centerline{$^1$ Dept. of Appl. Math., IPICyT, Apdo Postal 3-74 Tangamanga, San Luis Potos\'{\i}, MEXICO} 

\centerline{$^2$ Instituto de F\'{\i}sica, Universidad de Guanajuato, Apdo Postal E-143, Le\'on, MEXICO}

\bigskip



\bigskip

\noindent
{
{\bf Abstract} 

\noindent
We present several second-order linear differential equations that are associated to a particular Riccati equation with only one constant parameter in its
coefficients through the technique
of supersymmetric factorizations and through a Dirac-like procedure. The latter approach is a minimal extension of the results obtained with the first
technique in the sense that it includes up to two more constant parameters.  
\\}   
 
\bigskip


\bigskip


{\bf 1.} {\bf Riccati solutions and factorizations}

\bigskip

\noindent
We start with one of the simplest Riccati equation, which nevertheless dictates for example the behaviour of the Hubble 
constant of barotropic Friedmann-Robertson-Walker cosmologies \cite{far}
\begin{equation} \label{ricc}
{\rm u^{'}+cu^2+\kappa c=0~,}
\end{equation}
where $\kappa =\pm 1$ and ${\rm c}$ is a real constant.
Employing ${\rm u=\frac{1}{c}\frac{w^{'}_{\kappa}}{w_{\kappa}}}$ one gets the
well-known second order differential equation
\begin{equation} \label{schr1}
{\rm w^{''}_{\kappa}+\kappa c^2w_{\kappa}=0~.}
\end{equation}
For $\kappa =1$ the solution of the latter is
${\rm w_{1}=W_1\cos(c\eta +\varphi)}$, where $\varphi$ is an arbitrary phase,
whereas for $\kappa =-1$ one gets
${\rm w_{-1}=W_{-1}{\rm sinh}(c\eta)}$ 
where ${\rm W_{\pm 1}}$ 
are amplitude parameters.


The point now is that the Riccati solution
${\rm u_{p}=\frac{1}{c}\frac{w^{'}}{w}}$
mentioned above is only the particular solution, i.e.,
${\rm u_{p}=-\tan (c\eta )}$ and ${\rm u_{p}={\rm coth} (c\eta )}$
for $\kappa =\pm 1$, respectively. The particular Riccati solutions enter as nonoperatorial part in
the common factorizations of the second-order linear differential
equations that are directly related to the well-known Darboux isospectral transformations
\cite{ms}.
Indeed, Eq.~(\ref{schr1}) can be written
\begin{equation} \label{w}
{\rm w^{''}-c(-\kappa c)w=0}
\end{equation}
and also in factorized form using Eq.~(\ref{ricc}) one gets (${\rm D_{\eta}=\frac{d}{d\eta}}$)
\begin{equation} \label{w3}
{\rm \left(D_{\eta}+cu_{p}\right)
\left(D_{\eta}-cu_{p}\right)w=
w^{''}-c(u_{p}^{'}+cu_{p}^{2})w=0}~.
\end{equation}
To fix the ideas, we shall use the terminology of Witten's supersymmetric quantum mechanics and call Eq.~(\ref{w3}) the 
bosonic linear equation. On the other hand, the supersymmetric partner (or fermionic) 
equation of Eq.~(\ref{w3}) will be
\begin{equation} \label{f}
{\rm
\left(D_{\eta}-cu_{p}\right)
\left(D_{\eta}+cu_{p}\right)w_f=
w^{''}_{f}-c(-u_{p}^{'}+cu_{p}^2)w_{f}={\rm w^{''}_f
-c\cdot c_{\kappa,f}(\eta)w_f=0}}~,
\end{equation}
which is related to the fermionic Riccati equation
\begin{equation} \label{fR}
-{\rm u^{'}+cu^2-c_{\kappa,{\rm f}}(\eta)=0~.}
\end{equation}
Notice that for this fermionic Riccati equation the free term is already not constant.
Explicitly, one gets
$$
{\rm c_{\kappa,f}(\eta)=-u_{p}^{'}+cu_{p}^2=
\left\{ \begin{array}{ll}
{\rm c(1+2{\rm tan}^2 c\eta)} & \mbox{if $\kappa =1$}\\
{\rm c(-1+2{\rm coth}^2 c\eta)} & \mbox{if $\kappa =-1$}
\end{array} \right.}
$$
for the supersymmetric partner Riccati free term.
The solutions $\rm w_f$ are $\rm w_f =\frac{c}{\cos (c\eta +d)}$ 
and $\rm w_f =\frac{c}{sinh (c\eta)}$ for $\kappa =1$ and $\kappa =-1$,
respectively. 

Introducing the operator ${\rm P_{\eta}}=-i{\rm D_{\eta}}$ we can write the fermionic 
equations as follows
\begin{equation} \label{Pf}
{\rm
\left(-P_{\eta}-icu_{p}\right)
\left(P_{\eta}-icu_{p}\right)w_f=
-P_{\eta}^{2}w_{f}-c(-iP_{\eta}u_{p}+cu_{p}^2)w_{f}
}
~,
\end{equation}
whereas the bosonic case is
\begin{equation} \label{Pb}
{\rm
\left(P_{\eta}-icu_{p}\right)
\left(-P_{\eta}-icu_{p}\right)w_b=
-P_{\eta}^{2}w_{b}-c(iP_{\eta}u_{p}+cu_{p}^2)w_{b}
}
~,
\end{equation}

There is a more general bosonic factorization that has been introduced for the case of the quantum harmonic oscillator by Mielnik \cite{M84} that
in our case reads
 \begin{equation} \label{wbg}
{\rm \left(D_{\eta}+cu_{g,f}\right)
\left(D_{\eta}-cu_{g,f}\right)w_g=
w_g^{''}-c(u_{g,f}^{'}+cu_{g,f}^{2})w_g=w_g^{''}+\kappa cc(\eta ; \lambda)w_g=0}~.
\end{equation}
It is given in terms of the general Riccati solution  ${\rm u_{g,f}(\eta)}$ of the fermionic Riccati equation
(\ref{fR}) 
\begin{equation} \label{ug}
{\rm u_{g,f}}(\eta;\lambda)=  {\rm u_p(\eta)} -\frac{1}{{\rm c}} {\rm D_{\eta}}
\Big[ {\rm ln}({\rm I_{\kappa}}(\eta) + \lambda) \Big]
={\rm  D_{\eta}
\Big[ ln \left(\frac{w_{\kappa}(\eta)}{{\rm I_{\kappa}}(\eta) +
\lambda}\right)^{\frac{1}{c}}\Big]}
\end{equation}
and yields the one-parameter family of Riccati
free terms ${\rm c_{\kappa}(\eta;\lambda)}$ 
$$
{\rm -\kappa c_{\kappa}(\eta;\lambda)} = c{\rm u_{g}^2(\eta;\lambda) +
\frac{d u_{g}(\eta;\lambda)}{d\eta}}
= {\rm -\kappa c - \frac{2}{c}
D^2_{\eta} \Big[ ln({\rm  I}_{\kappa}(\eta) + \lambda)}
\Big]
$$
\begin{equation} \label{-kc3}
= {\rm -\kappa c - \frac{4 w_{\kappa}(\eta) w_{\kappa}^{\prime}
(\eta)}{c({\rm  I}_{\kappa}(\eta)
+ \lambda)}
+ \frac{2 w_{\kappa}^4(\eta)}{c({\rm  I}_{\kappa}(\eta) + \lambda)^2}~,}
\end{equation}
where ${\rm {\rm I}_{\kappa}(\eta)= \int _{0}^{\eta} \,
w_{\kappa}^2(y)\, dy}$,
if we think of a half line problem for which $\lambda$ is a positive
integration constant that occurs as a free parameter of the method.

The free terms ${\rm  c_{\kappa}(\eta;\lambda)}$ have the same
supersymmetric partner free term
${\rm c_{\kappa,f}(\eta).
}$ 
They may be considered
as intermediates between the initial constant Riccati free term ${\rm \kappa c}$ and
its supersymmetric partner
free term ${\rm c_{\kappa,f}(\eta)}$. 
From Eq.~(\ref{wbg}) one can infer the new parametric `zero mode' solutions of the 
linear equations entailing the functions
 ${\rm c_{\kappa}(\eta;\lambda)}$ as follows
\begin{equation} \label{wg}
{\rm w_g(\eta;\lambda)= 
\frac{w_{\kappa}(\eta)}{{\rm I}_{\kappa}(\eta) + \lambda}
}~.
\end{equation}

For $\lambda \rightarrow \infty$ the solutions ${\rm w_g}$ vanish. For an application to the damped classical 
oscillator see \cite{rore}. In quantum mechanics, where we have different types of solutions and physical interpretation, 
one can obtain ${\rm w_g}\rightarrow w_{\kappa}$ in the same 
asymptotic limit of the parameter if one introduces the appropriate normalization constant \cite{nc}. 


\newpage
\bigskip
{\bf 2.}  {\bf Dirac-like approach}

\bigskip

\noindent
The Dirac equation in the susy nonrelativistic formalism has been discussed by Cooper {\em et al} \cite{cooper} already in 1988.
They showed that the Dirac equation with a Lorentz scalar potential is associated with a susy pair of Schroedinger Hamiltonians.
This result has been used later by many authors in the particle physics context. In mathematical terms, it is only a simple approach for 
matrix differential equations.
Here we present several applications that may be considered the simplest extension of the results of the previous section in the sense 
that only up to two more constant parameters are introduced.   

{\bf 2.1} -
Let us now consider the following two Pauli matrices $\alpha =-{\rm i}\sigma _y=-{\rm i}\left( \begin{array}{cc}
0 & -{\rm i }\\
{\rm i} & 0\end{array} \right ) $ and $\beta =\sigma _x=\left( \begin{array}{cc}
0 & 1\\
1 & 0 \end{array} \right ) $ and write a Dirac-like equation for zero mass and at fixed zero energy of the form
\begin{equation} \label{HD}
D1\equiv {\rm  [i\sigma _y P_{\eta}+\sigma _x (icu_p)]}W=0~,
\end{equation}
where $W=\left( {\rm \begin{array}{cc}
w_1\\
w_2\end{array}} \right ) $ is a two component `zero-mass' spinor. 
This is equivalent to the following decoupled equations
\begin{eqnarray}
{\rm -P_{\eta}w_1+icu_pw_1=0}\\
+{\rm P_{\eta}w _2+icu_pw_2=0}~.
\end{eqnarray}
Solving these equations one gets ${\rm w}_1\propto 1/\cos ({\rm c}\eta)$ and ${\rm w}_2\propto \cos({\rm c}\eta)$ for the $\kappa =1$ case
and ${\rm w}_1\propto 1/{\rm sinh} ({\rm c}\eta)$ and ${\rm w}_2\propto {\rm sinh}({\rm c}\eta)$ for the $\kappa =-1$ case.
Thus, we obtain 
\begin{equation}\label{W1}
W=\left( {\rm \begin{array}{cc}
{\rm w_1}\\
{\rm w_2}\end{array}} \right )=\left( {\rm \begin{array}{cc}
{\rm w_f}\\
{\rm w_b}\end{array}} \right )~.
\end{equation}
This shows that the matrix `zero-mass' Dirac equation is equivalent to the two second-order linear differential  
equations of bosonic and fermionic type, Eq.~(\ref{schr1}) and Eq.~(\ref{f}), respectively \cite{cooper}.  

\bigskip

{\bf 2.2} -
Consider now the following Dirac equation 
\begin{equation} \label{HDM}
D2\equiv {\rm [i\sigma _y P_{\eta}+\sigma _x (icu_p +K)]W=KW}~,
\end{equation}
where $K$ is a positive real constant. In the left hand side, $K$ stands as a mass parameter of the Dirac spinor, whereas 
on the right hand side it corresponds to the energy parameter, i.e., $E=K$. Thus, we have a Dirac equation for a spinor
of mass $K$ at the fixed energy $E=K$. This equation  
can be written as the following system of coupled equations
\begin{eqnarray}
{\rm -P_{\eta}w _1+(icu_p+K)w _1=Kw _2}\\
{\rm P_{\eta}w _2+(icu_p+K)w _2=Kw _1}~.
\end{eqnarray}
This system is equivalent to the following second order equations for the two spinor components, respectively
\begin{equation} \label{sch1der}
{\rm -P_{\eta}^{2}w _{i}-c\Big[ i(\mp P_{\eta}-2K)u_p+
 cu_{p}^{2}\Big]w _{i}=0}~,
\end{equation}
where the subindex $i=1,2$.

The fermionic spinor components can be found directly as solutions of
\begin{equation} \label{comp1} 
{\rm D^{2}_{\eta}w_1^{+}-\Big[c^2(1+2\tan ^2 c\eta)+2icK\tan c\eta\Big] w_1^{+}=0  \qquad {\rm for} \, \kappa =1}
\end{equation}
and
\begin{equation} \label{comp1b} 
{\rm D^{2}_{\eta}w_1^{-}-\Big[c^2(-1+2{\rm coth}^2 c \eta)-2icK{\rm coth} \,c\eta\Big] w_1^{-}=0  \qquad  {\rm for} \, \kappa =-1}~,
\end{equation}
whereas the bosonic components are solutions of
\begin{equation} \label{comp2} 
{\rm D^{2}_{\eta}w_2^{+}+\Big[c^2-2icK\tan c\eta\Big] w_2^{+}=0  \qquad {\rm for} \quad  \kappa =1}
\end{equation}
and 
\begin{equation} \label{comp2b} 
{\rm D^{2}_{\eta}w_2^{-}+\Big[-c^2+2icK{\rm coth} \,c\eta\Big] w_2^{-}=0  \qquad {\rm for} \quad \kappa =-1}~.
\end{equation}
The solutions of the bosonic equations are expressed in terms of the Gauss hypergeometric functions $_2F_1$ in the variables ${\rm y=e^{ic\eta}}$ and
${\rm y=e^{c\eta}}$, respectively
$$
{\rm w_2^{+}=Ay^{-p}\, _{2}F_1\Big[-\frac{1}{2}(p+iq);-\frac{1}{2}(p-iq), 1-p; -y^2\Big]}+
$$
\begin{equation} \label{s1}
{\rm By^{p} \, _2F_1\Big[\frac{1}{2}(p-iq), \frac{1}{2}(p+iq),1+p; -y^2\Big]}
\end{equation}
and 
$$
{\rm w_2^{-}=C(-1)^{-\frac{i}{2}r}y^{-ir}\, _{2}F_1\Big[-\frac{i}{2}(r+is),-\frac{i}{2}(r-is), 1-ir; y^2\Big]}+ 
$$
\begin{equation} \label{s2}
{\rm D(-1)^{\frac{i}{2}r}y^{ir}\, _{2}F_1\Big[\frac{i}{2}(r-s),\frac{i}{2}(r+s), 1+ir; y^2\Big]}~,
\end{equation}
respectively. The parameters are the following: ${\rm p=(-1-\frac{2K}{c})^{\frac{1}{2}}}$, ${\rm q=(1-\frac{2K}{c})^{\frac{1}{2}}}$,
${\rm r=(-1-i\frac{2K}{c})^{\frac{1}{2}}}$, ${\rm s=(-1+i\frac{2K}{c})^{\frac{1}{2}}}$, whereas $\rm A$, $\rm B$, $\rm C$, $\rm D$ are superposition 
constants.

It is not necessary to try to find the general fermionic solutions through the analysis of their differential equations (\ref{comp1}) and (\ref{comp1b}) 
because they are related in a known way to the bosonic solutions \cite{ros}.
The general fermionic solutions can be obtained easily if one argues that the particular fermionic zero mode is the inverse of a particular
bosonic zero mode and constructing the other independent zero mode solution as in textbooks. Thus
\begin{equation}\label{s3}
{\rm w_1}^{\pm}=\frac{1 +k\int ^{y}[{\rm w_2}^{\pm}]^2dz}{{\rm w_2}^{\pm}}~,
\end{equation}
where $k$ is an arbitrary constant.

\bigskip

{\bf 2.3} -
The most general case in this scheme with only constant parameters is to consider the following matrix Dirac-like equation
$$
D3\equiv\Bigg[{\rm i}\left( \begin{array}{cc}
0 & -{\rm i }\\
{\rm i} & 0\end{array} \right ){\rm P_{\eta}}+\left( \begin{array}{cc}
0 & 1\\
1 & 0 \end{array} \right )\left( \begin{array}{cc}
 {\rm icu_p +K_1}& 0\\
0 &{\rm  icu_g+K_2}\end{array} \right )\Bigg]\left( \begin{array}{cc}
{\rm w}_1\\
{\rm w}_2 \end{array} \right )=
$$
\begin{equation} \label{Dg}
\left( \begin{array}{cc}
{\rm K_1}& 0\\
0 &{\rm  K_2}\end{array} \right )\left( \begin{array}{cc}
{\rm w_1}\\
{\rm w_2} \end{array} \right )~.
\end{equation}
Proceeding as in {\bf 2.2} one finds the coupled system of first-order differential equations
\begin{eqnarray}
\Big[{\rm P_{\eta}}+{\rm icu_g}+{\rm K_2}\Big]{\rm w_2}={\rm K_1}{\rm w_1}\\
\Big[-{\rm P_{\eta}}+{\rm ic u_p}+{\rm K_1}\Big]{\rm w_1}={\rm K_2}{\rm w_2}
\end{eqnarray}
and the equivalent second-order differential equations
$$
-{\rm P_{\eta}}^{2}{\rm w} _{i}+\Big[{\rm ic(u_p-u_g)}+({\rm K_1-K_2})\Big]{\rm P_{\eta}}{\rm w}_i+
$$
\begin{equation} \label{Schrg}
\Big[ {\rm ic}(\pm {\rm P_{\eta}}{\rm u}_i+{\rm K_1}{\rm u_g}+{\rm K_2}{\rm u_p})
 -{\rm c^2}{\rm u_p u_g}\Big]{\rm w} _{i}=0~,
\end{equation}
where the subindex $i=1,2$, and ${\rm u_1}$ and ${\rm u_2}$ correspond to ${\rm u_p}$ and ${\rm u_g}$, respectively.
In the ${\rm D_{\eta}}$ notation this equation reads
$$
{\rm D_{\eta}}^{2}{\rm w} _{i}+\Big[{\rm c}\Delta {\rm u_{pg}}-{\rm i}\Delta {\rm K})\Big]{\rm D_{\eta}}{\rm w}_i+
$$
\begin{equation} \label{Schrgb}
+\Big[{\rm c}(\pm {\rm D_{\eta}}{\rm u}_i+({\rm iK}_1{\rm u_g}+{\rm K_2}{\rm u_p}))
 -{\rm c}^2{\rm u_{p}u_g}\Big]{\rm w} _{i}=0~.
\end{equation}
Under the gauge transformation 
\begin{equation} \label{gauge}
{\rm w} _{i}=z_{i}\exp \left(-\frac{1}{2}\int ^{\eta}\Big[{\rm c}\Delta {\rm u_{pg}}-{\rm i}\Delta {\rm K})\Big]d\tau\right)=z_i(\eta)\frac{e^{\frac{1}{2}{\rm i}\eta \Delta {\rm K}}}{({\rm I_{\kappa}+\lambda})^{\frac{1}{2}}}\end{equation}
one gets
\begin{equation} \label{schz}
-{\rm P_{\eta}}^{2}z_i+Q_i(\eta)z_i=0,  \quad {\rm or} \quad {\rm D_{\eta}}^{2}z_i+Q_i(\eta)z_i=0,
\end{equation}
where 
$$
Q_i(\eta)=\Big[{\rm c}(\pm {\rm D_{\eta}u}_i+({\rm iK_1}{\rm u_g}+{\rm K_2}{\rm u_p}))
 -{\rm c}^2{\rm u_{p}u_g}\Big]-
$$
\begin{equation} \label{Q}
\frac{1}{2}{\rm D_{\eta}}\Big[{\rm c}\Delta {\rm u_{pg}}\Big]-\frac{1}{4}\Big[{\rm c}\Delta {\rm u_{pg}}-{\rm i}\Delta {\rm K})\Big]^2
\end{equation}
for $i=1,2$, respectively. Since $Q_i$ are complicated functions we were not able to find analytical solutions of Eq.~(\ref{schz}).

The corresponding Dirac spinor is of the following form
\begin{equation} \label{ww3}
W(\eta;{\rm K_1,K_2}, \lambda)=\left( {\rm \begin{array}{cc}
z_1(\eta ;{\rm K_1})\\
z_2(\eta ;{\rm  K_2}, \lambda)\end{array}} \right )=\left( {\rm \begin{array}{cc}
 {\rm w}_{\rm f}(\eta; {\rm K_1})\\
{\rm w}_{\rm g}(\eta; {\rm K_2}, \lambda)\end{array}} \right )~,
\end{equation}
where $ {\rm w}_{\rm g}(\eta; {\rm K_2},\lambda)$ is given by Eq.~(\ref{wg}) when ${\rm K_1=K_2=0}$. The discussion of the asymptotic 
limit $\lambda \rightarrow \infty$ is similar to that at the end of the first section.

\bigskip


\bigskip

\end{document}